\documentclass[twocolumn,showpacs,amsmath,amstex,amssymb,mathfonts,prl]{revtex4-1}
\usepackage{graphicx,bm,units}
\usepackage{cases}
\usepackage{enumerate}
\newcommand{\beq}{\begin{equation}}\newcommand{\eeq}{\end{equation}}\newcommand{\beqa}{\begin{eqnarray}}
\newcommand{\eeqa}{\end{eqnarray}}\newcommand{\w}{\wedge}
\newcommand{\dl}{\bm{\delta}}\newcommand{\nn}{\nonumber}

\begin{document}
\title{Torsional Monopoles and Torqued Geometries in Gravity and Condensed Matter}

\author{Andrew Randono$^1$}\email{arandono@perimeterinstitute.ca}\author{Taylor L. Hughes$^2$}
\affiliation{$^{1}$The Perimeter Institute for Theoretical Physics, 31 Caroline St. North,Waterloo, ON N2L 2Y5, Canada}
\affiliation{$^2$Department of Physics, University of Illinois, 1110 West Green St, Urbana IL 61801} 

\begin{abstract}
Torsional degrees of freedom play an important role in modern gravity theories as well as in condensed matter systems where they can be modeled by defects in solids. Here we isolate a class of torsion models that support torsion configurations with a localized, conserved charge that adopts integer values. The charge is topological in nature and the torsional configurations can be thought of as torsional `monopole' solutions. We explore some of the properties of these configurations in gravity models with non-vanishing curvature, and discuss the possible existence of such monopoles in condensed matter systems. To conclude, we show how the monopoles can be thought of as a natural generalization of the Cartan spiral staircase.
\end{abstract}
\maketitle
The analogy between geometry and defects in gravity theories and in theories of elasticity in solids is an old and well developed field of study \cite{Ruggiero:2003tw,Katanaev:2004xq,Lazar:2001hs,Hehl:2007bn,kleinertBook}. Disclinations and dislocations in crystals are defects in the ordered lattice which carry finite curvature and torsion respectively. Transporting a particle around a disclination (dislocation) produces a non-zero rotation (translation) by the end of the cycle. Dislocations are particularly interesting because, while sources of curvature are ubiquitous in the natural universe, the effects of torsion in a gravitational context are thus far negligible experimentally\cite{hammond2002}. In solids, however, dislocations affect many important material properties and are present even in the cleanest materials. Thus condensed matter systems can provide useful laboratories for the study of torsion. 

\begin{figure}
 \begin{center}
\includegraphics[height=7.0cm]{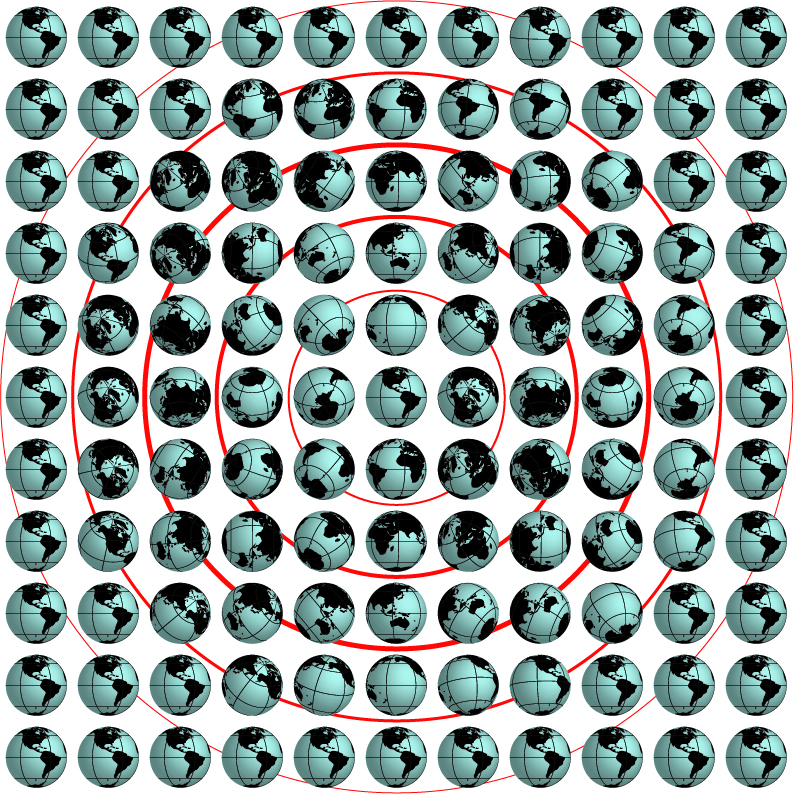}
   \end{center}
  \caption{\label{MonopoleWorlds}A cross-section through the origin of a torsion monopole with  $Q=1$. The globe picks out three directions which form an orthonormal triad. The monopole pictured has a radius of five lattice sites, and it is formed by rotating each globe along a radial line directed from the origin by an amount proportional to the radius of the lattice site, until the angle of rotation reaches $2\pi$ at $r=5$.
  }
 \end{figure}

As we will show, the defects that we will describe cannot be described by the classical geometric theory of elasticity. Instead, these defects may occur in materials described by micropolar elasticity theory\cite{eringen1967}. Micropolar elasticity theory (or Cosserat elasticity) is a simple extension of classical elasticity to include local orientational degrees of freedom of the constituent particles/molecules of the elastic medium. The defect we investigate, which we dub a torsional `monopole' (TM), is a defect that does not require a lattice deformation, but a deformation texture in the local rotational degrees of freedom. Such defects could exist in biological or granular systems (two common systems described by micropolar elasticity) and may affect solids with a strong coupling between orbital electronic motion and local spin or orbital degrees of freedom. Although TMs are not very complicated objects, as evidenced by the structure shown in Fig. \ref{MonopoleWorlds}, there are some subtle issues that require a careful treatment. We present a general treatment of these defects in a gravitational context in flat and curved space and then give an explicit construction of 
a TM  and its relation to defects in solids and the ``Cartan spiral staircase."
We begin with a simple formulation of our construction in flat space. To isolate the purely torsional degrees of freedom in the absence of other geometric constructs, we will begin with three highly constraining ans\"{a}tze. In order to focus on the minimal, kinematical properties of torsional defects, and to make the theory as generalizable as possible to various condensed matter and gravity theories, we will not assume an underlying dynamical geometric theory. More specifically, we will begin with the typical ingredients of Einstein-Cartan gravity, but we will not impose  the Einstein-Cartan or any other dynamical equations of motion. For generality we will work with the 4D Lorentzian theory, but all of the major results are applicable to 3D Euclidean systems. Thus, we take the geometry to be described by a $Spin(3,1)\simeq \overline{SO}(3,1)\simeq SL(2,\mathbb{C})$ gauge theory whose connection coefficient in a local trivialization is the spin-connection $\omega$, and a (co)-frame (tetrad) field $e$, which is a one-form taking values in the adjoint representation of $Spin(3,1)$. 
The three ans\"{a}tze we will make are
\begin{enumerate}[i. ]
\item{The manifold has the topology of $\mathbb{R}^4$ and the metric induced from the co-frame $e^I$ is the flat Minkowski metric \emph{i.e.}  in Cartesian coordinates 
$\bm{g}=\eta_{IJ} \,e^I\otimes e^J =-dt^2+dx^2+dy^2+dz^2.$} 
\item{The spin connection $\omega^{IJ}$ has vanishing curvature; thus,  we have a Weitzenb\"{o}ck spacetime.}
\item{The spin connection asymptotically approaches the Levi-Civita connection (the special connection compatible with $e^I$) at spatial infinity so that $\lim_{r\rightarrow \infty}\omega = \Gamma[e]$.}
\end{enumerate}
We refer to a geometry satisfying these conditions as a {\it torqued geometry} for reasons that will become clear. 

We will denote the Levi-Civita connection associated with the tetrad by $\Gamma=\Gamma[e]$. For any connection $A$ let the curvature $R_{A}=dA+A\wedge A.$ Compatibility means the torsion associated with this connection is zero $D_{\Gamma}e^I =de^I+\Gamma^I{}_J\w e^J=0.$
Since the metric is Minkowski,  $R_{\Gamma}=0.$ Additionally, the assumption that the space is Weitzenb\"{o}ck implies $R_{\omega}=0$ as well. 
However, we are not assuming that the spin-connection is \emph{compatible} with the tetrad, and the (generically non-zero) torsion is given by the usual expression (in the adjoint representation)
\beq
T^I=D_\omega e^I =de^I+\omega^I{}_J\w e^J\,. 
\eeq
To simplify the statement of the third ansatz we express the spin-connection $\omega$ in terms of the Levi-Civita connection $\Gamma$ by
\beq
\omega^{I}{}_J=\Gamma^{I}{}_J+ C^{I}{}_J \label{Relation0}
\eeq
where $C^{IJ}=C^{[IJ]}{}_\mu \,dx^\mu$ is known as the contorsion tensor. This can always be carried out since the space of connections is an affine space. The torsion is simply
\beq
T^I=C^I{}_J\w e^J
\eeq\noindent and the assumption that the spin-connection asymptotically approaches the Levi-Civita connection at spatial infinity implies that  $\lim_{r\rightarrow \infty}T^I=0.$ The \emph{rate} at which the torsion must tend to zero will be fixed later.

In Minkowski space all flat connections are gauge related (since $\pi_1(\mathbb{R}^4)=0$) and thus, we can express the spin-connection in terms of the Levi-Civita connection as (in the fundamental representation, where we will drop explicit indices)
\beq
\omega=g\Gamma g^{-1} -dg\,g^{-1}\,. \label{Relation1}
\eeq
The group element $g=g(x)$ is the ``relative gauge" between the spin connection and the tetrad. The term {\it torqued geometry} is in reference to the relationship between the spin connection and  Levi-Civita connection in Eq. \ref{Relation1}. Using Eqs. (\ref{Relation0}) and (\ref{Relation1})  we have:
\beq
C=g \Gamma g^{-1} -dg\,g^{-1} -\Gamma=-D_{\Gamma}g \,g^{-1}\,.\label{ContorsionA}
\eeq
We will occasionally use the `trivial' gauge where the tetrad in Cartesian coordinates is  $\overset{0}{e}{}^I=\delta^I_\mu \,dx^\mu$, and the corresponding Levi-Civita connection $\Gamma[\overset{0}{e}]=0$. For $\overset{0}{e}=heh^{-1}$ the spin connection is
\beqa
\omega'=h \omega h^{-1} -dh h^{-1}=-dg' g'^{-1}  \quad \text{with} \;\; g'=hgh^{-1}\,. \label{SpinG}
\eeqa

 One of the key properties of the torsional configurations that follows from our flatness ansatz is the existence of a conserved current. To see this, consider the curvature of the spin connection expressed in terms of the contorsion tensor. From the definition we have
\beqa
R_{\omega}&=& R_\Gamma +D_\Gamma C +C \w C \nn\\ 
&=& D_{\Gamma}C +C\w C=0
\eeqa
which follows since both $R_\Gamma,R_\omega=0.$ 
Now consider 
\beqa
\Omega &\equiv& Tr_D\left(\frac{1}{4\pi^2}(D_{\Gamma}C+C\w C)\w (D_{\Gamma}C+C\w C)\right) \nonumber\\
&=& \frac{1}{4\pi^2}d\;Tr_D\left(C \w D_\Gamma C+\dfrac{2}{3} C\w C\w C\right)\equiv d\xi^{(3)}_C
\eeqa\noindent with $Tr_D(\cdot)=\frac{1}{D}Tr(\cdot)$ where $D$ is the dimension of the representation.
Defining the (Hodge-dual of the) current by $*J\equiv \xi^{(3)}_{C}$ and recognizing that $\Omega =0$, we see that the current is conserved: $d*\!J=0.$
We can also define an analogous ``dual" current from the identity
\beqa
\Omega_\star &\equiv& Tr_D\left(\frac{1}{4\pi^2}\star (D_{\Gamma}C+C\w C)\w (D_{\Gamma}C+C\w C)\right) \nn\\
&=& d\left(\frac{1}{4\pi^2}Tr_D\left(\star C \w D_\Gamma C+\dfrac{2}{3} \star C\w C\w C\right)\right)\,,
\eeqa
where $\star$ is the Hodge-dual operator in the internal $Spin(3,1)$ vector space.

Both of these currents allow us to define a conserved charge. To do this, we will first fix our asymptotic boundary conditions on the torsion so that the defects we will consider are spatially isolated. It is sufficient to assume that in the trivial gauge,  $C,$ and thus $T,$ fall off like $\frac{1}{r}$ as $r \rightarrow \infty .$  Thus, in this gauge the relative gauge given in Eq. (\ref{SpinG}), must be such that $g' \rightarrow constant$ near spatial infinity. This allows for the standard compactification $\Sigma \simeq  \mathbb{R}^3 \cup \{\infty\} \simeq \mathbb{S}^3$ for a spatial slice $\Sigma.$ Due to the assumed fall-off conditions on the torsion, the flux of current through the timelike cylinder at asymptotic infinity is zero, so 
we can therefore define the {\it conserved} torsional-charge 
\beqa
Q&\equiv &\int_\Sigma \xi^{(3)}_{C}= -\frac{1}{12\pi^2}\int_\Sigma Tr_D\left( C \w C\w C\right)\,. \label{Charge}
\eeqa
Similarly the dual torsional charge is given by
\beq
Q_\star \equiv  \frac{1}{4\pi^2}\int_\Sigma Tr_D\left( \star \,C \w D_\Gamma C+2/3 \star C\w C\w C\right).\nonumber
\eeq  The conservation of these charges is a purely kinematic property, \emph{independent} of any dynamics to which the TMs are subjected.  Indeed, the charges  are \emph{topological} in nature and under small deformations  $\{\dl e, \dl \omega\}$ that preserve the flatness constraints we have $\dl Q=0$ and $\dl Q_\star=0.$ 
In fact, $Q$ takes quantized integer values (as we will now show), whereas the charge $Q_\star$ is identically zero. 

First, we note that despite its similarity to the Chern-Simons functional, the charge $Q$ is different in that it is identically gauge \emph{invariant} under both large and small gauge transformations. Thus, we can choose a convenient gauge in order to compute the charge. We choose the trivial gauge where the contorsion is given by $C=-dg' g'^{-1}$ (c.f. Eq. \ref{SpinG}).  Thus in this gauge using Eq. (\ref{Charge}), we have (using the shorthand notation $(dg g^{-1})^3 =Tr_D(dg g^{-1}\w dg g^{-1}\w dg g^{-1})$)
\beqa
Q&=& \frac{1}{12 \pi^2} \int_\Sigma (dg' g'^{-1})^3 = \frac{1}{12 \pi^2} \int_\Sigma (dg g^{-1})^3\,.
\eeqa
We recognize the last line as the index or winding number of the relative gauge viewed as a map $g:\Sigma \rightarrow Spin(3,1)$. Despite  $Spin(3,1)$ being non-compact, the winding number is well defined since $Spin(3,1)$ has $SU(2)$ as its maximal compact subgroup. Such maps are classified by $\pi_3(Spin(3,1))=\mathbb{Z}.$ We conclude that $Q$ is conserved and takes integer values, so $Q\in \mathbb{Z}.$

 It is a simple matter to construct explicit configurations with non-zero charge. To do this, we borrow from well-known results in $SU(2)$ Yang-Mills theories (see e.g. \cite{Rajaraman:Instantons}). We will work in the trivial gauge in Cartesian coordinates so the tetrad is $e^I_\mu=\delta^I_\mu$ and $\Gamma^{IJ}=0$. In this gauge $\omega=-dg' g'^{-1}$, and defining $\tau_i =\frac{1}{2}\epsilon_{ijk}\gamma^j \gamma^k$, where $\gamma^I$ are generators of the Clifford algebra,  and $i,j,k...=1,2,3...$ are spatial indices, we take $g'$ to be ($(i)$ labels which TM)
\beqa
g'=g_{(i)}=\cos(\chi_{(i)}) \,\mathbf{1} + \sin(\chi_{(i)}) \, \frac{x^a-x^a_{(i)}}{|\vec{x}-\vec{x}_{(i)}|}\,\tau_a \label{g}
\eeqa
where $\chi_{(i)}=\chi_{(i)}(\Delta r_{(i)})$ with $\Delta r_{(i)}=|\vec{x}-\vec{x}_{(i)}|$ is any continuous and differentiable function that monotonically increases from $0$ at $\Delta r_{(i)}=0$ to $\pi$ at $\Delta r_{(i)}=\infty$. To ensure that the configuration is well behaved, we will also assume that $\frac{\partial \chi_{(i)}}{\partial \Delta r_{(i)}} \big|_{\Delta r_{(i)}=0}=\frac{\partial \chi_{(i)}}{\partial \Delta r_{(i)}} \big|_{\Delta r_{(i)}=\infty}=0$.
For a single TM of charge $q$ located at the origin $(\vec{x}_{(i)}=0)$, the contorsion is given explicitly by
\beqa
C^{ij}=&-&2\left[\,{\epsilon^{ij}}_k\,\hat{X}^k d\chi +\sin(\chi) \cos(\chi)\,{\epsilon^{ij}}_k \,d\hat{X}^k\right.\nonumber\\
&-&\left. 2\sin^2(\chi) \,\hat{X}^{[i} d\hat{X}^{j]}  \right]. \label{Contorsion1}
\eeqa
The torsional charge for this configuration can be explicitly computed to yield $Q=1$.
We can then use this group element to generate multiple TM solutions of the generic form 
\beqa
C= -dg' g'^{-1} \quad \quad \text{with} \quad \quad g'=g^{q_1}_{(1)} g_{(2)}^{q_2} \dots g_{(N)}^{q_N} \label{Monopoles}
\eeqa\noindent and charge  $Q=q_1+q_2+ \dots +q_N.$

It is worthwhile to address a potential source of confusion stemming from the analogous geometric constructs in Yang-Mills theory. We have referred to the configurations above as monopoles because they are spatially isolated torsional defects of a topological nature. However, typical nomenclature in Yang-Mills theories associates  monopoles with $\pi_2(G)$, and instantons with $\pi_3(G)$. Despite similarities to analogous structures in Yang-Mills theories, the TM has some fundamental differences. The key property that allows for a stable, gauge invariant topological structure in three dimensions is that the topological charge can be identified not with a single Chern-Simons functional but with the difference of two Chern-Simons functionals: $Q=\int_{\Sigma}(CS(\omega)-CS(\Gamma))$. The resulting quantity is invariant under large gauge transformations, unlike either of its two constituents, but the quantity picks out the winding number of the relative gauge between the two connections.

 Interestingly, our major results can be extended to curved spacetime.  The key property we want to retain is the existence of a conserved current with an associated topologically quantized charge. The key is to preserve the fundamental property of a torqued geometry. That is, given a tetrad $e$ and its associated Levi-Civita connection $\Gamma$, the spin connection differs only by a relative gauge:
\beq
\omega=g\Gamma g^{-1} -dg\,g^{-1} \quad \quad \quad R_{\omega}=g R_{\Gamma} g^{-1} \,.
\label{CurvedR}
\eeq
The contorsion is still given Eq. \ref{ContorsionA}
but the curvature is no longer zero so generically the contorsion satisfies the condition
$D_{\Gamma}C+C\w C=g R_{\Gamma}g^{-1}-R_{\Gamma}. $
Nevertheless, there is still a conserved current. To see this, consider the four-form (which is zero from Eq. (\ref{CurvedR}))
\beq
\Omega=\frac{1}{4\pi^2}Tr_{D}\left( R_{\omega}\w R_{\omega} -R_{\Gamma}\w R_{\Gamma}\right) =0\,.
\eeq
This gives rise to the conserved current $*J$ since
\beqa
\Omega&=& d (CS(\omega)-CS(\Gamma))\equiv  d * J =0 \,.
\eeqa
The current is entirely torsional in nature as it can be written
\beq
*J=\xi^{(3)}_{C}+\frac{1}{2\pi^2} \,Tr_D \left( 2\, C\w R_{\Gamma} \right)\,.
\eeq
This current is conserved and being the difference of two Chern-Simons functionals that differ by a relative gauge, $g$, we clearly have $Q=\frac{1}{12 \pi^2} \int_{\mathbb{S}^3}(dg\,g^{-1})^3.$
Thus, the charge is topologically quantized. One can also construct the ``dual" charge $Q_\star$, but this charge vanishes.

Now we return to flat space to discuss the analogy with defects in solids.  Thus far, we have viewed torqued geometries as a deformation of the spin-connection by a relative gauge transformation. To model a TM it is convenient to make a (true, not relative) gauge transformation to absorb the deformation entirely in the tetrad. It is sufficient to work with 3d Euclidean space and we denote the triad by $E^i_a.$ The geometric variables describing a torqued geometry before the transformation are $
E=\overset{0}{E},$ and $\omega=g \overset{0}{\Gamma} g^{-1} -dg \,g^{-1}$
where $\overset{0}{E}$ is a fiducial Euclidean flat tetrad and $\overset{0}{\Gamma}=\Gamma[\overset{0}{E}]$ is the corresponding Levi-Civita connection. Gauge transforming by $g^{-1}$ we obtain
$ E'=g^{-1} \overset{0}{E} g$ and $\omega'=g^{-1}\omega g- dg^{-1}\, g=\overset{0}{\Gamma}.$
Now the deformation induced by the relative gauge is encoded in the triad as opposed to the spin connection. For a single TM of charge $q$ located at the origin, the relative gauge $g=\cos(q \,\chi(r)) \mathbf{1}+ i\sin(q\,\chi(r)) \,\hat{x}_a\, \overset{0}{E}{}^a_i \,\sigma^i$ where $\sigma^i$ are the Pauli matrices is an element of $SU(2)$ at each point. For clarity, fix the fiducial triad to be  $\overset{0}{E}{}^i_a=\delta^i_a$. We focus on the behavior of $E^{'i}_a$ along any ray beginning at the origin and ending at asymptotic infinity. The gauge transformation represents a rotation at each point around the axis defined by the ray. Since $\chi(0)=0$  at the origin, and monotonically increases to $\chi(\infty)=\pi, $ the relative gauge represents a spatial rotation of the fiducial triad around the radial axis such that the rotation is the identity at the origin and increases monotonically to $2\pi q$ at infinity. The direction is clockwise or counterclockwise depending on the sign of $q$. 

The realization of a model for the TM in a condensed matter system points out some of the deficiencies of the classical geometric theory of elasticity. The classical geometric theory of dislocations and disclinations has  
 the Euclidean Poincar\'{e} group $G=SO(3)\ltimes T(3)$ as a gauge group where $T(3)$ is the set of three dimensional translations \cite{Mermin:1979zz}.  Disclinations are defects associated with the rotational degrees of freedom, and in this model, the absence of disclinations is associated with the vanishing of the curvature of the spin connection $R_{\omega}=0$ \cite{Ruggiero:2003tw,Katanaev:2004xq,Lazar:2001hs,Hehl:2007bn,kleinertBook}. Thus, the TM, which can exist in geometries with vanishing curvature is not composed of disclinations. Next we consider pure dislocations which are described by a non-zero torsion associated with the translation group. In a solid we have a lattice which breaks the continuous translation symmetry down to a discrete subgroup. The topological charges of defects in the translation sector are thus given by $\pi_{n}(T(3)/\mathbb{Z}\oplus \mathbb{Z}\oplus \mathbb{Z})=\pi_{n}(T^3)$ where $\pi_n$ is the $n$-th homotopy group, $\mathbb{Z}\oplus \mathbb{Z}\oplus \mathbb{Z}$ represents the space of 3d discrete translations,  and $T^3$ is the 3-torus. The only topologically stable defects are line defects (dislocations due to $\pi_1(T^3)=\mathbb{Z}\oplus \mathbb{Z}\oplus \mathbb{Z}$) and thus our \emph{point-like} TM is not a dislocation. In fact, these arguments are immediately apparent if we choose the gauge  for the TM where the entire deformation is in the triad. This deformation does not require a lattice deformation and thus is not effectively captured in classical elasticity theory. 

To support the TM we need consider the local rotational degrees of freedom of the objects forming the elastic medium and thus materials described by micropolar elasticity\cite{eringen1967,Hehl:2007bn}.  We imagine molecules or grains to which a local triad is associated. This set of axes describes the local orientation of the molecule. A fiducial geometry has all the local triads aligned and the TM is a defect texture centered the origin (see Fig. \ref{MonopoleWorlds}).  The molecules along a radial ray are rotated around the ray-axis. If the orientations are realigned outside a radius $R$ then the topological charge is equal to the number of revolutions carried out  between $r=0$ and $r=R.$ 
 In the continuum theory, the gauge group (in the 3D Euclidean case) is $G=SU(2)$, and the isometry group is the subgroup that leaves the triad fixed, namely the rotations by $2\pi$ forming a $\mathbb{Z}_2$ subgroup. Thus, the relevant homotopy group is $\pi_3(SU(2)/\mathbb{Z}_2)=\pi_3(SO(3))$. The existence of the TM is a reflection of the property $\pi_3(SO(3))=\mathbb{Z}.$ It is likely that these defects will affect the elastic behavior of micropolar media, but perhaps it is more interesting to consider possible effects in the electronic behavior of solids. Although we leave this analysis open for future work we comment that these defects would likely affect the electronic behavior of materials with strong spin-orbit coupling since they are very sensitive to the local orientation of the orbitals. In fact, the 3D massive Dirac Hamiltonian, the minimal model for a topological insulator\cite{hasan2010,qi2010}, is effectively minimally coupled to the triad field and will be affected by a TM. 

The TM we have described also provides a generalization of Cartan's spiral staircase to the spherical case (which we might refer to as Cartan's spiral stairway to heaven). Cartan's spiral staircase is the name given to a simple model for a space with torsion first described in 1922 \cite{Cartan:Spiral,Hehl:2007bn}. For this construction, we choose a particular  $\chi(r)$ that makes the analogy as clear as possible. Suppose one were sitting on the surface of a sphere (say the earth) at radius $r_0$ and desired to build a spiral stairway analogous to Cartan's spiral staircase, but oriented in the radial direction. We will build the stairway in discrete steps to illustrate the point. As opposed to the original spiral staircase, our model is easiest to understand by rotating the Euclidean triad and fixing the spin connection to be $\omega^{ij}=0$, as we saw in the previous section. First take the triad  $E'$ above with 
\beq
\chi(r)=
\begin{cases}
0 & \text{for   } 0\leq r \leq r_0 \\
\frac{\pi}{\lambda} (r-r_0) & \text{for   } r_0 \leq r \leq r_0+\lambda  \\
\pi & \text{for   } r_0+\lambda \leq r \,.
\end{cases}
\eeq
This gives a TM localized within radius $r_0+\lambda$ with torsional charge $Q=1$. Using the geometric description above, this TM is easy to visualize. Traveling from the sphere at $r_0$ outward to  $r_0+\lambda$ the triad rotates by one full turn around the radial ray. This gives the TM, and the first stage of the construction of Cartan's spherical spiral stairway. To extend the stairway to $r_0+2\lambda$ we use a the radial function
\beq
\chi(r)=
\begin{cases}
0 & \text{for   } 0\leq r \leq r_0 \\
\frac{\pi}{\lambda} (r-r_0) & \text{for   } r_0 \leq r \leq r_0+2\lambda  \\
2\pi & \text{for   } r_0+2\lambda \leq r \,.
\end{cases}
\eeq
This now describes a TM with charge $Q=2$ that extends to $r_0+2\lambda$, and adds one more turn to the spiral stairway. We can keep extending the stairway turn by turn such that at the $q$-th stage of construction, the staircase extends to $r_0+q\lambda$ and the number of turns (and the TM charge, $Q$) is $q$. Carrying out this procedure {\it ad infinitum} as $q\rightarrow \infty$ generates Cartan's spiral stairway to heaven.

{\it Acknowledgements} We thank F. W. Hehl and R. G. Leigh for helpful comments. AR was supported by the NSF International Research Fellowship Grant \#0853116. TLH was supported by NSF DMR 0758462 and the ICMT at UIUC. TLH thanks the Perimeter Institute for generous hosting when this work was initiated. 


\begin{thebibliography}{10}%
\makeatletter
\providecommand \@ifxundefined [1]{%
 \ifx #1\undefined \expandafter \@firstoftwo
 \else \expandafter \@secondoftwo
\fi
}%
\providecommand \@ifnum [1]{%
 \ifnum #1\expandafter \@firstoftwo
 \else \expandafter \@secondoftwo
\fi
}%
\providecommand \enquote [1]{``#1''}%
\providecommand \bibnamefont  [1]{#1}%
\providecommand \bibfnamefont [1]{#1}%
\providecommand \citenamefont [1]{#1}%
\providecommand\href[0]{\@sanitize\@href}%
\providecommand\@href[1]{\endgroup\@@startlink{#1}\endgroup\@@href}%
\providecommand\@@href[1]{#1\@@endlink}%
\providecommand \@sanitize [0]{\begingroup\catcode`\&12\catcode`\#12\relax}%
\@ifxundefined \pdfoutput {\@firstoftwo}{%
 \@ifnum{\z@=\pdfoutput}{\@firstoftwo}{\@secondoftwo}%
}{%
 \providecommand\@@startlink[1]{\leavevmode}%
 \providecommand\@@endlink[0]{}%
}{%
 \providecommand\@@startlink[1]{%
  \leavevmode
  \pdfstartlink
   attr{/Border[0 0 1 ]/H/I/C[0 1 1]}%
   user{/Subtype/Link/A<</Type/Action/S/URI/URI(#1)>>}%
  \relax
 }%
 \providecommand\@@endlink[0]{\pdfendlink}%
}%
\providecommand \url  [0]{\begingroup\@sanitize \@url }%
\providecommand \@url [1]{\endgroup\@href {#1}{\urlprefix}}%
\providecommand \urlprefix [0]{URL }%
\providecommand \Eprint[0]{\href }%
\@ifxundefined \urlstyle {%
  \providecommand \doi [1]{doi:\discretionary{}{}{}#1}%
}{%
  \providecommand \doi [0]{doi:\discretionary{}{}{}\begingroup
  \urlstyle{rm}\Url }%
}%
\providecommand \doibase [0]{http://dx.doi.org/}%
\providecommand \Doi[1]{\href{\doibase#1}}%
\providecommand \bibAnnote [3]{%
  \BibitemShut{#1}%
  \begin{quotation}\noindent
    \textsc{Key:}\ #2\\\textsc{Annotation:}\ #3%
  \end{quotation}%
}%
\providecommand \bibAnnoteFile [2]{%
  \IfFileExists{#2}{\bibAnnote {#1} {#2} {\input{#2}}}{}%
}%
\providecommand \typeout [0]{\immediate \write \m@ne }%
\providecommand \selectlanguage [0]{\@gobble}%
\providecommand \bibinfo [0]{\@secondoftwo}%
\providecommand \bibfield [0]{\@secondoftwo}%
\providecommand \translation [1]{[#1]}%
\providecommand \BibitemOpen[0]{}%
\providecommand \bibitemStop [0]{}%
\providecommand \bibitemNoStop [0]{.\EOS\space}%
\providecommand \EOS [0]{\spacefactor3000\relax}%
\providecommand \BibitemShut [1]{\csname bibitem#1\endcsname}%
\bibitem{Ruggiero:2003tw}%
  \BibitemOpen
  \bibfield{author}{%
  \bibinfo {author} {\bibfnamefont{M.~L.}\ \bibnamefont{Ruggiero}}\ and\
  \bibinfo {author} {\bibfnamefont{A.}~\bibnamefont{Tartaglia}},\ }%
  \bibfield{journal}{%
  \bibinfo {journal} {Am. J. Phys.}\ }%
  \textbf{\bibinfo {volume} {71}},\ \bibinfo {pages} {1303} (\bibinfo {year}
  {2003})%
  \bibAnnoteFile{NoStop}{Ruggiero:2003tw}%
\bibitem{Katanaev:2004xq}%
  \BibitemOpen
  \bibfield{author}{%
  \bibinfo {author} {\bibfnamefont{M.~O.}\ \bibnamefont{Katanaev}},\ }%
  \bibfield{journal}{%
  \bibinfo {journal} {Phys. Usp.}\ }%
  \textbf{\bibinfo {volume} {48}},\ \bibinfo {pages} {675} (\bibinfo {year}
  {2005})%
  \bibAnnoteFile{NoStop}{Katanaev:2004xq}%
\bibitem{Lazar:2001hs}%
  \BibitemOpen
  \bibfield{author}{%
  \bibinfo {author} {\bibfnamefont{M.}~\bibnamefont{Lazar}},\ }%
  \bibfield{journal}{%
  \bibinfo {journal} {J. Phys.}\ }%
  \textbf{\bibinfo {volume} {A35}},\ \bibinfo {pages} {1983} (\bibinfo {year}
  {2002})%
  \bibAnnoteFile{NoStop}{Lazar:2001hs}%
\bibitem{Hehl:2007bn}%
  \BibitemOpen
  \bibfield{author}{%
  \bibinfo {author} {\bibfnamefont{F.~W.}\ \bibnamefont{Hehl}}\ and\ \bibinfo
  {author} {\bibfnamefont{Y.~N.}\ \bibnamefont{Obukhov}}}%
   (\bibinfo {year} {2007}),\
  \Eprint{http://arxiv.org/abs/0711.1535}{arXiv:0711.1535 [gr-qc]}%
  \bibAnnoteFile{NoStop}{Hehl:2007bn}%
\bibitem{kleinertBook}%
  \BibitemOpen
  \bibfield{author}{%
  \bibinfo {author} {\bibfnamefont{H.}~\bibnamefont{Kleinert}},\ }%
  \emph{\bibinfo {title} {Multivalued Fields: In Condensed Matter,
  Electromagnetism, and Gravitation}},\ \bibinfo {edition} {2008th}\ ed.\
  (\bibinfo {publisher} {World Scientific},\ \bibinfo {year} {2008})%
  \bibAnnoteFile{NoStop}{kleinertBook}%
\bibitem{hammond2002}%
  \BibitemOpen
  \bibfield{author}{%
  \bibinfo {author} {\bibfnamefont{R.~T.}\ \bibnamefont{Hammond}},\ }%
  \bibfield{journal}{%
  \bibinfo {journal} {Rep. Prog. Phys.},\ \bibinfo {pages} {599}}%
   (\bibinfo {year} {2002})%
  \bibAnnoteFile{NoStop}{hammond2002}%
\bibitem{eringen1967}%
  \BibitemOpen
  \bibfield{author}{%
  \bibinfo {author} {\bibfnamefont{A.~C.}\ \bibnamefont{Eringen}},\ }%
  \bibfield{journal}{%
  \bibinfo {journal} {Int. J. Eng. Sci.},\ \bibinfo {pages} {191}}%
   (\bibinfo {year} {1967})%
  \bibAnnoteFile{NoStop}{eringen1967}%
\bibitem{Rajaraman:Instantons}%
  \BibitemOpen
  \bibfield{author}{%
  \bibinfo {author} {\bibfnamefont{R.}~\bibnamefont{Rajaraman}},\ }%
  \emph{\bibinfo {title} {Solitons and Instantons: an introduction to solitons
  and instantons in Quantum Field Theory}},\ \bibinfo {edition} {1989th}\ ed.\
  (\bibinfo {publisher} {North Holland},\ \bibinfo {year} {1989})%
  \bibAnnoteFile{NoStop}{Rajaraman:Instantons}%
\bibitem{Mermin:1979zz}%
  \BibitemOpen
  \bibfield{author}{%
  \bibinfo {author} {\bibfnamefont{N.~D.}\ \bibnamefont{Mermin}},\ }%
  \bibfield{journal}{%
  \Doi{10.1103/RevModPhys.51.591}{\bibinfo {journal} {Rev. Mod. Phys.}}\ }%
  \textbf{\bibinfo {volume} {51}},\ \bibinfo {pages} {591} (\bibinfo {year}
  {1979})%
  \bibAnnoteFile{NoStop}{Mermin:1979zz}%
\bibitem{hasan2010}%
  \BibitemOpen
  \bibfield{author}{%
  \bibinfo {author} {\bibfnamefont{M.~Z.}\ \bibnamefont{Hasan}}\ and\ \bibinfo
  {author} {\bibfnamefont{C.~L.}\ \bibnamefont{Kane}},\ }%
  \bibinfo {howpublished} {arxiv: 1002.3895}%
  \bibAnnoteFile{NoStop}{hasan2010}%
\bibitem{qi2010}%
  \BibitemOpen
  \bibfield{author}{%
  \bibinfo {author} {\bibfnamefont{X.-L.}\ \bibnamefont{Qi}}\ and\ \bibinfo
  {author} {\bibfnamefont{S.~C.}\ \bibnamefont{Zhang}},\ }%
  \bibinfo {howpublished} {arxiv: 1008.2026}%
  \bibAnnoteFile{NoStop}{qi2010}%
\bibitem{Cartan:Spiral}%
  \BibitemOpen
  \bibfield{author}{%
  \bibinfo {author} {\bibnamefont{\'{E}lie Cartan}},\ }%
  \bibfield{journal}{%
  \bibinfo {journal} {C.R. Acad. Sci.},\ \bibinfo {pages} {593}}%
   (\bibinfo {year} {1922})%
  \bibAnnoteFile{NoStop}{Cartan:Spiral}%
\end{thebibliography}
\end{document}